\documentclass[conference]{IEEEtran}
\IEEEoverridecommandlockouts

\usepackage{cite}
\usepackage{amsmath,amssymb,amsfonts}
\usepackage{algorithmic}
\usepackage{graphicx}
\usepackage{textcomp}
\usepackage{xcolor}
\usepackage{graphicx}
\usepackage{booktabs}
\usepackage{url}
\usepackage{multirow}
\usepackage{threeparttable}

\usepackage[most]{tcolorbox}

\newcommand{\censortext}[1]{\textcolor{black}{\rule{3em}{1ex}}}

\def\BibTeX{{\rm B\kern-.05em{\sc i\kern-.025em b}\kern-.08em
    3T\kern-.1667em\lower.7ex\hbox{E}\kern-.125emX}}
    
\begin{document}
\title{On Categorizing Open Source Software Security Vulnerability Reporting Mechanisms on GitHub}

\author{%
Sushawapak Kancharoendee\textsuperscript{\dag}, Thanat Phichitphanphong\textsuperscript{\dag}, Chanikarn Jongyingyos\textsuperscript{\dag}, Brittany Reid\textsuperscript{*},\\
Raula Gaikovina Kula\textsuperscript{\S}, Morakot Choetkiertikul\textsuperscript{\dag}, Chaiyong Ragkhitwetsagul\textsuperscript{\dag}, Thanwadee Sunetnanta\textsuperscript{\dag}\\
\textsuperscript{\dag}\textit{Faculty of Information and Communication Technology, Mahidol University, Thailand}\\
\textsuperscript{*}\textit{Graduate School of Science and Technology, Nara Institute of Science and Technology (NAIST), Japan}\\
\textsuperscript{\S}\textit{Graduate School of Information Science and Technology, Osaka University, Japan}\\
\{sushawapak.kan, thanat.phi, chanikarn.jon\}@student.mahidol.ac.th, brittany.reid@naist.ac.jp, \\raula-k@ist.osaka-u.ac.jp, \{morakot.cho, chaiyong.rag, thanwadee.sun\}@mahidol.ac.th
}

\maketitle

\begin{abstract}

Open-source projects are essential to software development, but publicly disclosing vulnerabilities without fixes increases the risk of exploitation. The Open Source Security Foundation (OpenSSF) addresses this issue by promoting robust security policies to enhance project security. Current research reveals that many projects perform poorly on OpenSSF criteria, indicating a need for stronger security practices and underscoring the value of SECURITY.md files for structured vulnerability reporting. This study aims to provide recommendations for improving security policies. By examining 679 open-source projects, we find that email is still the main source of reporting. Furthermore, we find that projects without SECURITY.md files tend to be less secure (lower OpenSSF scores). Our analysis also indicates that, although many maintainers encourage private reporting methods, some contributors continue to disclose vulnerabilities publicly, bypassing established protocols. The results from this preliminary study pave the way for understanding how developers react and communicate a potential security threat. Future challenges include understanding the impact and effectiveness of these mechanisms and what factors may influence how the security threat is addressed.

\end{abstract}


\section{Introduction}
\label{1_introduction}



Open-source software has become essential in modern development, enabling developers to reuse and modify code. However, with its growing use, the risk of vulnerabilities has increased \cite{Alfadel:EMSE2023}. As security threats rise, developers' security practices are essential \cite{Zahan:ICSE2023}. Effective management is key, especially for open-source projects, where developers' responses to vulnerabilities are vital \cite{Prana:EMSE2021}. Publicly disclosing vulnerabilities without fixes heightens the risk of exploitation. To address this, platforms like GitHub provide security measures, such as a \texttt{SECURITY.md} file \cite{GitHubSecurityPolicy:online} for reporting vulnerabilities and GitHub Advisories \cite{GitHubAdvisoryDatabase:online} for documenting security issues. 

\begin{figure}[!]
    \centering
    \includegraphics[width=0.95\linewidth]{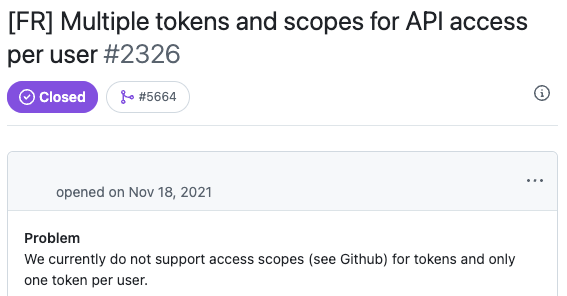}
    \caption{A security issue created in open source project}
    
    \label{fig:example_issue}
    \vspace*{-.75cm}
\end{figure}

To address bugs or discuss their projects, developers use GitHub Issues, a task-tracking feature available to all GitHub users for open-source projects. However, the traditional practice of reporting security-related issues through GitHub Issues can pose risks of vulnerability being known by attackers, as it makes vulnerabilities publicly known before they are fixed, giving attackers more time to exploit them. For this reason, some GitHub projects request that developers avoid reporting vulnerabilities through Issues. For example, Figure \ref{fig:example_issue} shows a security issue in an open-source project posted publicly as a GitHub issue, detailing a problem with token handling. By publically posting this issue, the unaddressed token leak could allow attackers to perform unauthorized actions beyond the intended scope before the developers can fix the issue.

Recent studies showed that only a small proportion of open-source projects have defined security policies \cite{Ayala:SVM2023}. However, defining security policy in the projects is considered an important practice for developers to improve the project's security \cite{Zahan:ICSE2023}. These highlight a gap in the security practices of the open-source projects. However, no study has focused on analyzing the characteristics and effectiveness of the security policies, including the extent to which developers' practices align with the provided guidelines.
The related work in this study examines security policies in open-source projects on GitHub. Ayala et al. \cite{Ayala:SVM2023} analyzed popular GitHub repositories, finding that only a small portion had implemented security policies, with many lacking structured security measures. This highlights the importance of workflows and security policies in open-source projects. The GitHub security policy feature guides vulnerability reporting, while security issues serve as a feature for bug tracking and project discussions. Bühlmann et al. \cite{Noah:SAC2022} studied security issues, noting that they comprise a small yet growing portion of issues and require more discussion time than non-security issues.
Zahan et al. \cite{Zahan:IEEE} used the OpenSSF Scorecard to assess security practices in the npm and PyPI ecosystems. The study identified gaps in areas such as Code Review, Maintenance, Licensing, Branch Protection, and Security Policy. The study also recommended practices for enhancing such areas. Additionally, a relationship between security practices and vulnerability levels was identified \cite{Zahan:ICSE2023}.

Despite increased attention on security policies for open-source projects, no research has yet examined the actual content of these policies. Our study addresses this gap by analyzing specific elements within security policies to propose guidelines for improvement. By identifying practices that could improve OpenSSF Scorecard scores if properly implemented, we set the stage for further guidance on defining effective security policies.

This motivated us to investigate the characteristics of security policies in open-source projects, aiming to provide software practitioners with guidelines for enhancing security policies. This paper presents our preliminary investigation into GitHub projects' security policies, focusing on two main aspects. First, we examine how developers report vulnerabilities by comparing the practices outlined in security policies with actual developer actions. To do this, we analyze the content of security policies to identify vulnerability reporting processes and manually review developer practices. Second, we conduct a preliminary assessment of security practices' quality using the Open Source Security Foundation's (OpenSSF) Scorecard \cite{Scorecard:online}. Through a comparative study, we evaluate security practices in projects with and without defined security policies to determine whether the presence of a security policy strengthens overall security practices. Thus, we define the following research questions for our preliminary investigation:

\begin{enumerate}
    \item \textbf{RQ1:} What are the reporting mechanisms in security policies?
    \item \textbf{RQ2:} Do the developer's practices align with the security policy?
    \item \textbf{RQ3:} Do projects with a security policy differ in OpenSSF Scorecard scores compared to those without one?
\end{enumerate}

To address our research questions, we analyzed 679 PyPi packages hosted on GitHub. We focused on PyPI ecosystem for Python language since Python is a widely used programming language in software development, as indicated in the 2024 IEEE Spectrum survey \cite{TopProgrammingLanguage:online}. We examined how developers report security vulnerabilities by reviewing security policy instructions and analyzing security issues in these projects. We then compared Scorecard scores between projects with and without a defined security policy. Our study shows that most PyPi package maintainers recognize the need for private channels like email or GitHub advisories to report vulnerabilities, as outlined in many security policies. However, external contributors often bypass these policies and submit security issues publicly. Projects with established security policies show notable improvements in the Scorecard's Dependency Update Tool and Maintained Practices scores. This underscores the importance of improving public communication strategies or refining security policies to enhance developer awareness and adherence to security practices.

 \section{Background}
\label{2_background}




This section covers background and related work, starting with an overview of GitHub security policies and the OpenSSF Scorecard for evaluating security practices in open-source projects.

\subsection{Security Policies in GitHub Repositories}
GitHub provides a feature for creating a security policy within a repository through a dedicated \texttt{SECURITY.md} file. This file guides software practitioners, such as contributors and users, on handling security concerns, including the appropriate channels for reporting vulnerabilities in a project. Maintainers can add the security policy to the repository’s root, \texttt{doc}, or \texttt{.github} folder. When creating a security policy using GitHub’s security feature, a guideline suggests including key details to improve its usefulness.\footnote{\url{https://docs.github.com/en/code-security/getting-started/adding-a-security-policy-to-your-repository}} These details include the \textit{Supported Versions} section, listing the project versions that receive security updates, and the \textit{Reporting a Vulnerability} section, which provides instructions for reporting vulnerabilities. This section can also specify communication channels for updates on reported vulnerabilities, outline steps taken upon vulnerability acceptance or rejection, and include additional information about the repository’s security practices. 

GitHub also provides features that enable users to report vulnerabilities, through the GitHub Advisory (GHSA) \cite{GitHubAdvisoryDatabase:online}. GitHub Advisory is a database that publishes information on security vulnerabilities that impact open-source projects. It includes security-related details, such as severity, affected versions, and recommended fixes.



\subsection{OpenSSF Scorecard}
The OpenSSF Scorecard is an automated tool developed by the Open Source Security Foundation (OpenSSF) to assess security practices in open-source software repositories \cite{OpenSSF:online}. The Scorecard provides a security practices score for each project, based on 18 criteria that collectively reflect the robustness of security practices in maintaining a project repository. These scores help software practitioners identify and address potential security risks in their projects, offering insights to improve overall security practices. 


Each metric in the OpenSSF Scorecard is classified into one of four risk categories: Critical, High, Medium, and Low, based on its potential security implications. Metrics are scored from 0 to 10, reflecting how closely the project aligns with security best practices. If Scorecard cannot verify a particular practice, it assigns a score of -1, which indicates inconclusive evidence or a possible runtime error. This designation ensures that such cases do not negatively impact the overall project score. For example, the Scorecard evaluates metrics like Branch Protection, which checks if protective measures are in place for critical branches, and Dependency Update Tool, assessing whether tools are used to keep dependencies up-to-date, which helps reduce vulnerability risks. Another metric, License, confirms whether the project has a well-defined, open-source license, contributing to transparency and legal security. These checks help developers understand security gaps and guide improvements.

\begin{figure*}[ht!]
    \centering
    \includegraphics[width=0.85\linewidth]{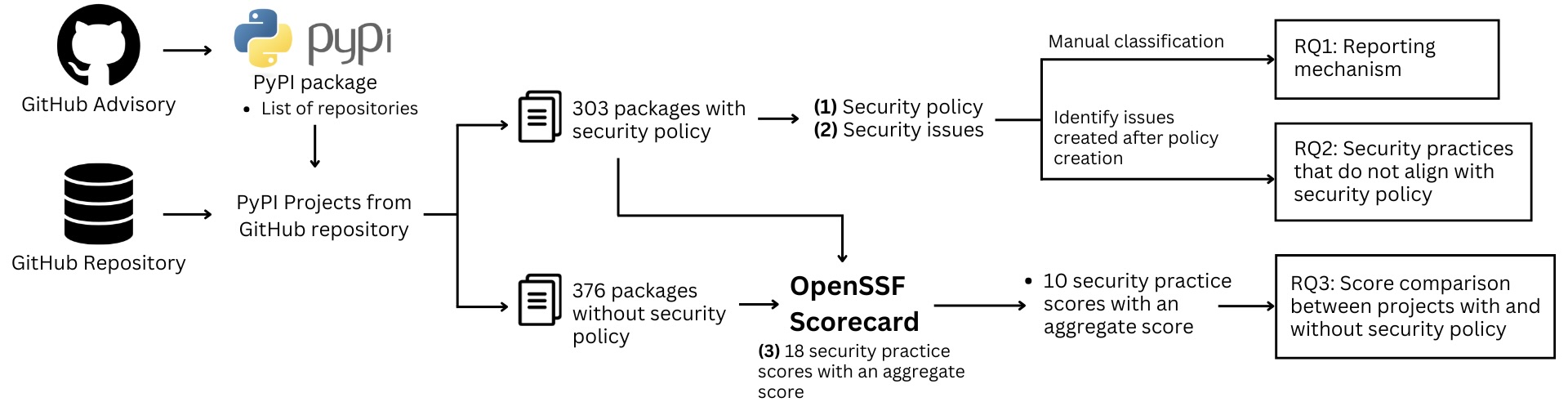}
    \caption{Our research methodology}
    \label{fig:workflow}
\end{figure*}


\section{Methodology}
\label{3_methodology}
In this section, we explain our methodology, detailing the data collection process and our approach to answering the research questions.

\subsection{Overview}
Figure \ref{fig:workflow} shows an overview of the study workflow. To answer research questions, we collect PyPI packages from the security advisories in the GitHub advisory database. Our methodology examines how developers report vulnerabilities as outlined in security policies and compares this with developers' actual practices. We manually categorize the instructions provided in these policies and then identify any security issues that do not conform to them. To highlight the importance of effective security policies, we assess security practices using the OpenSSF Scorecard for both packages with and without a security policy. By comparing the scores, we identify specific practices that show significant improvements when a security policy is implemented.




\subsection{Data collection}
Our study incorporates three data sets: (1) GitHub repositories of the PyPi packages that appear in the GitHub advisory database, (2) security issues from GitHub repositories, and (3) security practice scores from the OpenSSF Scorecard. We gathered GitHub repositories of the PyPi packages with at least one GitHub advisory record from GitHub Advisory Database \cite{GitHubAdvisoryDatabase:online}, which includes vulnerability advisories published before May 2024. The initial dataset contains 2,982 security advisories for PyPi packages. To ensure the accuracy of the collection process, we verified key repository values of the advisory, \texttt{source\_code\_location}, which links to the reported repositories.


In the GitHub security-related issues collection, we utilized WordNet \cite{WordNetDoc:online} to identify synonyms and related words based on a core set of terms: [``vulnerability'', ``security'', ``risk'', ``CVE'', ``CWE'']. We collected the issues that have the labels containing these security-related keywords. The dataset of security issues includes the metadata of the issues, which comprises links to the issues, titles, labels with security keywords, submitter information, and submission dates. Lastly, we used the OpenSSF Scorecard tool to assess the security practices of collected repositories. The dataset includes scores for 18 security practices along with an overall security score, with each practice score ranging from -1 to 10.


\subsection{Data processing}
We began by performing data filtering to ensure the quality and relevance of the repositories analyzed. First, we excluded the OpenSSF Scorecard criterion rated as -1, as this rating indicates that the tool was unable to generate a score due to insufficient data or an internal error during the assessment process. Additionally, we excluded one repository that contained over 1,000 issue reports, treating it as an outlier due to its atypically high volume of issues. Following this initial filtering, we categorized the repositories into two groups. The first group included repositories with a defined security policy, identified by the presence of \texttt{SECURITY.md}, \texttt{security.md}, or \texttt{security.rst} files in either the top-level directory or the \texttt{.github} folder. 

The second group comprised repositories without any security policy files. This classification allowed us to compare security practices between projects with and without explicit security policies in place. Note that the OpenSSF Scorecard considers the presence of a security policy under the ``Security Policy'' criterion. To ensure a fair comparison, we excluded the ``Security Policy'' score from our analysis, as the existence of a security policy directly impacts the overall OpenSSF score. In conclusion, our dataset comprises 679 repositories, with 303 containing the security policy and 376 without one.







\subsection{Data Analysis}
To address RQ1, we analyze the content specified in security policies. GitHub security policies typically guide developers on reporting vulnerabilities and security issues, detailing the reporting mechanisms available. The classification process was conducted collaboratively among the authors. We manually classified these mechanisms by examining the content to identify the specific methods or channels provided for issue reporting one by one. When discovering new reporting mechanisms, we expanded the methods and conducted the classification iteratively. To ensure consistency and minimize the influence of individual biases, we calculated Cohen’s Kappa \cite{jacob_cohen:1960} for the reliability of the manual analysis, achieving an agreement score of 0.9. Our analysis revealed four distinct reporting mechanisms: reporting through email, external links (e.g., proprietary issue tracking systems), GitHub advisories, and GitHub Issue creation. The definition and example practices are presented in Table \ref{tab:reporting_mechanism_def}. This classification allowed us to identify the most commonly used reporting methods.

\begin{table}[h!]
\centering
\caption{Reporting Mechanism Definition}
\label{tab:reporting_mechanism_def}
\begin{tabular}{p{1.2cm}p{3.2cm}p{2.7cm}}
\toprule
\textbf{Reporting Mechanism} & \textbf{Definition} & \textbf{Example} \\ \midrule
Email & The project provides the process to report the vulnerabilities through sending an email & If you discover a security vulnerability or would like to report a security issue privately and securely, please email us at \censortext{security@openmicroscopy.org.} \\

GitHub Advisory & The project provides the process to report the vulnerabilities by submitting a GitHub Advisory & To report a vulnerability head over to the Security Advisories page and click on "New draft security advisory". \\

GitHub Issue & The project provides the process to report the vulnerabilities by creating an issue & Please report a vulnerability as an issue; our team will evaluate it and address it in the proper time. \\

External link & The project provides another platform (e.g., bug bounty program) or external links for further information about the security policy & Please report sensitive security issues via \censortext{Spotify's} bug-bounty program by following this instruction, rather than GitHub. \\

Not mentioned & The project did not provide the reporting mechanism in the security policy & \\ \bottomrule

\end{tabular}
\end{table}

For RQ2, we examine repositories that specify reporting methods such as Email, External Links, and GitHub Advisory in their security policies. Specifically, we assess whether, after the declaration of these policies, any security-related issues were reported as GitHub Issues, which would be non-compliant with the stated reporting methods. 

To address RQ3, we perform a statistical comparison of OpenSSF Scorecard scores between repositories with a security policy and those without one.

\section{Results}
\label{4_results}

This section provides a detailed account of the results for each research question, along with a discussion that emphasizes the significance of our findings.



\subsection{RQ1: What are the reporting mechanisms in security policies?}



\begin{figure}[!]
    \centering
    \includegraphics[width=0.95\linewidth]{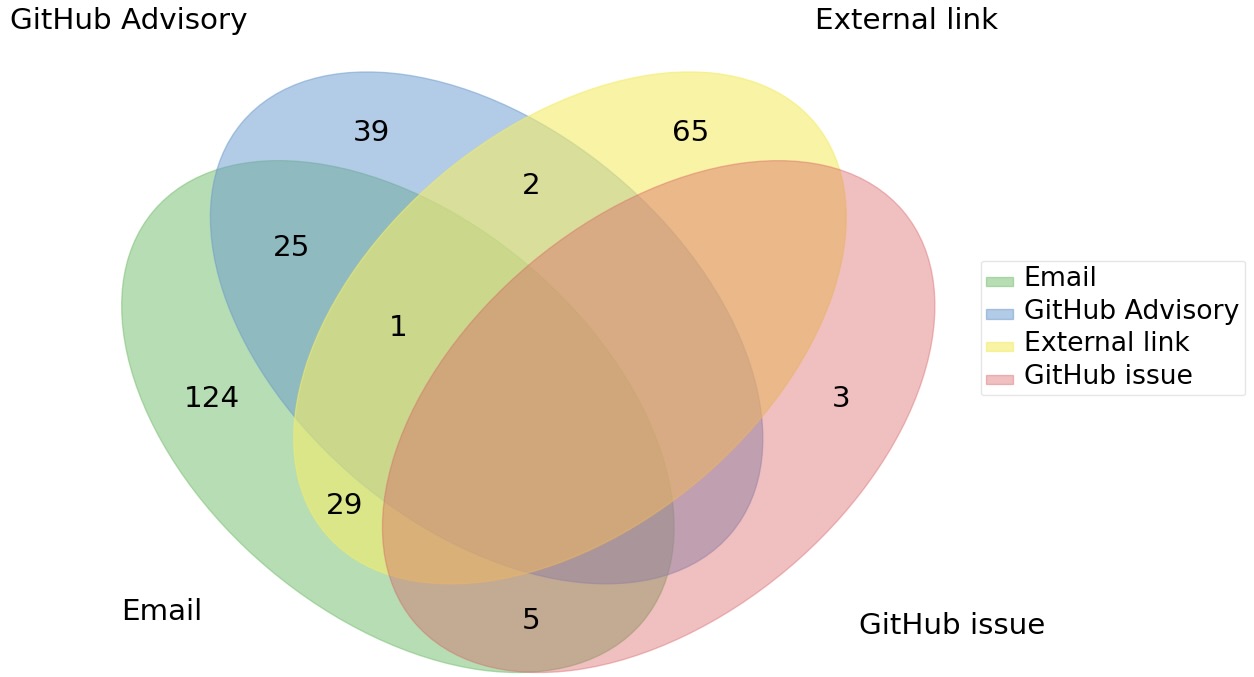}
    \caption{The Reporting Mechanism Defined In Security Policy}
    \label{fig:rq1}
    \vspace*{-.5cm}
\end{figure}

Figure \ref{fig:rq1} shows a Venn diagram of the reporting mechanisms defined in security policies. Our analysis of 302 GitHub repositories from PyPi shows that most repositories use email (41.06\%) as the preferred reporting method, followed by external links to other platforms (21.52\%) and the GitHub Advisories feature (12.91\%). Notably, 62 repositories (20.53\%) provide more than one reporting mechanism. Only 0.99\% use GitHub Issues, and 2.98\% do not specify any reporting method. We categorize email, external links, or GitHub advisories as private channels for reporting security vulnerabilities in repositories. In answering RQ1, we found that most security policies define the private channels as reporting mechanisms, as the results show that 285 repositories (94.37\%) define these channels in the security policies. This suggests that most project maintainers are aware of the risks of publicly disclosed security vulnerabilities and take steps to mitigate them.

\textbf{RQ1 Summary:} 
Our findings indicate that most project maintainers are aware of publicly disclosed security, since most security policy reporting mechanisms are private communication channels.

\subsection{RQ2: Do the developer's practices align with the security policy?}

We investigate the alignment between reporting mechanisms defined in security policies and actual practices. In repositories specifying Email, External Link, or GitHub Advisory as reporting channels, we found a significant decrease in security-related issues reported through GitHub Issues, suggesting that practitioners are generally adhering to the defined reporting mechanisms. However, 787 security-related issues were still reported through GitHub Issues across 58 repositories after private reporting methods were established, indicating non-compliance with the policies. 

Table \ref{tab:issue_reporting_mechaism_submitter} shows that of 787 non-compliance issues, these are primarily occurring in repositories using Email as the reporting method. 
We also investigated the roles of submitters of these security issues identified by the \texttt{author\_association} attribute of the analyzed repositories. This attribute includes four categories: \textit{OWNER}, \textit{CONTRIBUTOR}, and \textit{MEMBER}, while \textit{NONE} represents external contributors with no formal association to the project. 
The results indicate that most developers follow the defined reporting mechanism, as the results show a small number of security issues do not comply with the security policy. However, for these noncompliant security issues, the majority of them were created by external contributors (i.e., the \texttt{author\_association} is \textit{NONE}), who may either be unaware of the policy or intentionally disregard it.

\begin{table}[h!]
\centering
\caption{Non-compliance in security-related issues based on reporting mechanisms and reporter roles}
\label{tab:issue_reporting_mechaism_submitter}
\begin{tabular}{p{2.9cm}r p{2.0cm}r}
\toprule
\textbf{Reporting mechanism} & \textbf{\%} & \textbf{Submitter} & \textbf{\%} \\ \midrule
Email & 39.52 & NONE & 44.34 \\
External link & 26.30 & MEMBER & 25.67 \\
GitHub (GH) Advisory & 21.09 & CONTRIBUTOR & 23.13 \\
Email and External link & 9.66 & COLLABORATOR & 6.73 \\
Not mentioned & 1.78 & OWNER & 0.13 \\
Email \& GH Advisory & 0.76 &  &  \\
Email \& GH issues & 0.76 &  &  \\ 
GH Advisory \& External link & 0.13 &  &  \\ \bottomrule
\end{tabular}
\end{table}


 \textbf{RQ2 Summary:} 
Most developers follow the defined reporting mechanism, as a small number of security issues that do not comply with the policy are created by external contributors.




\subsection{RQ3: Do projects with a security policy differ in OpenSSF Scorecard scores compared to those without one?}

We used the OpenSSF Scorecard tool to assess 302 repositories with a security policy and 376 without one across 10 measurable security practices from OpenSSF Scorecard: Binary Artifacts, Branch Protection, CII Best Practices, Contributors, Dependency Update Tool, Fuzzing, License, Maintained, SAST, and Vulnerabilities. 

Table \ref{tab:SSF_score_statistics} shows the comparison results. Repositories with a security policy in place show higher adherence to security practices across most metrics. Notably, the mean scores for eight practices such as Branch Protection, CII Best Practices, and Maintained, are significantly higher in repositories with a security policy. This is shown by a higher mean score with \textit{p-value} less than 0.001 for the repositories with the security policy. This indicates that repositories with security policies are more proactive about implementing strong security practices.

\textbf{RQ3 Summary:} Repositories with security policies are more proactive in implementing security practices, as evidenced by the higher security practices score.

\begin{table}[h!]
\centering
\caption{Comparison of OpenSSF Scorecard scores between repositories with and without security policies}
\label{tab:SSF_score_statistics}
\begin{tabular}{@{}lrrrrrrr@{}}
\toprule
\multirow{2}{*}{\textbf{Security Practice}} & \multicolumn{2}{c}{\textbf{With Policy}} & \multicolumn{2}{c}{\textbf{Without Policy}} & \multirow{2}{*}{\textbf{p-value}} \\ \cmidrule(lr){2-3} \cmidrule(lr){4-5}
 & \textbf{Mean} & \textbf{SD} & \textbf{Mean} & \textbf{SD} &  \\ \midrule
\textbf{Aggregate Score*} & 5.93 & 1.14 & 3.95 & 1.29 & $<$0.001 \\
\textbf{Binary-Artifacts} & 9.50 & 1.73 & 9.64 & 1.61 & 0.26 \\
\textbf{Branch-Protection*} & 3.53 & 2.94 & 1.69 & 2.57 & $<$0.001 \\
\textbf{CII-Best-Practices*} & 0.29  & 1.21 & 0.01 & 0.1 & $<$0.001 \\
\textbf{Contributors*} & 9.48 & 1.97 & 8.25 & 3.37 & $<$0.001 \\
\textbf{Dependency-Update-Tool*} & 7.02 & 4.58 & 3.19 & 4.67 & $<$0.001 \\
\textbf{Fuzzing*} & 1.75 & 3.81 & 0.61 & 2.40 & $<$0.001 \\
\textbf{License} & 9.45 & 1.71 & 9.24 & 2.29 & 0.20 \\
\textbf{Maintained*} & 7.49 & 4.12 & 4.04 & 4.62 & $<$0.001 \\
\textbf{SAST*} & 3.57 & 4.44 & 1.06 & 2.84 & $<$0.001 \\
\textbf{Vulnerabilities} & 7.45 & 3.9 & 7.43 & 3.95 & 0.95 \\ \bottomrule
\end{tabular}
\centering
\begin{tablenotes}
\item{* indicates practices where the score for repositories with a security policy is significantly higher than those without.}
\end{tablenotes}

\end{table}

\section{Challengs and Future Outlook}
\label{6_conclusion}

Open-source projects play a vital role in software development, but sharing vulnerabilities before fixes are available poses a significant risk of exploitation. Consequently, open-source projects often establish their own security policies to address these concerns. Our study examined security reporting mechanisms in PyPi packages on GitHub, focusing on the alignment between security policies and actual reporting practices. Interestingly, email remains the primary means of reporting security issues. Our analysis reveals frequent non-compliance in reporting security issues, suggesting that current reporting methods may not be effective for all contributors.
Effective security policies must cover key practices such as secure coding, vulnerability management, data handling, and access control. A crucial takeaway is that email communication requires robust security measures to protect sensitive information. The OpenSSF Scorecard analysis also shows that projects with well-defined security policies tend to adhere more closely to key practices, underscoring the positive impact of effective security policies on project security. At this stage, we observe various methods of reporting vulnerabilities, which warrants further investigation into the reasons behind these choices and their pros and cons.

Future work could focus on understanding the communities of open-source projects and their practices and compliance that they enforce within their communities. This includes examining software library ecosystems, where regulations in an open-source context remain unclear. Examining security policy and identifying best practices for security policies across diverse open-source ecosystems is also needed. This would involve tailoring security policy elements to varying project characteristics, such as project size, complexity, and contributor base, to provide actionable recommendations for software practitioners. 

Moreover, in the era of AI, further research is needed to explore automated methods for detecting non-compliance with security policies and investigate communication strategies to enhance adherence among external contributors, such as tools for preventing vulnerabilities from being discussed publicly. Such agents would be critical in quickly raising awareness and mitigating potential threats while identifying false positives. Ultimately, these challenges and future directions offer new opportunities to keep the open-source community secure from security threats with rapid mitigation.

\section{Acknowledgment}
\label{7_acknowledgment}

This work is supported by JSPS KAKENHI JP20H05706, JP23K28065, and Daiichi-Sankyo ``Habataku" Support Program for the Next Generation of Researchers, NAIST Senju Monju Project.

\bibliographystyle{ieeetr}
\bibliography{bibliography}

\end{document}